# Trust Evaluation Using an Improved Context Similarity Measurement

Mohsen Raeesi[1], Mohammad Amin Morid [2], Mehdi Shajari[3]

Department of Computer Engineering and Information Technology,
Amirkabir University of Technology, Iran

## Abstract

*In context-aware trust evaluation, using ontology tree is a popular approach to represent the relation between contexts. Usually, similarity between two contexts is computed using these trees. Therefore, the performance of trust evaluation highly depends on the quality of ontology trees. Fairness or granularity consistency is one of the major limitations affecting the quality of ontology tree. This limitation refers to inequality of semantic similarity in the most ontology trees. In other words, semantic similarity of every two adjacent nodes is unequal in these trees. It deteriorates the performance of contexts similarity computation. We overcome this limitation by weighting tree edges based on their semantic similarity. Weight of each edge is computed using Normalized Similarity Score (NSS) method. This method is based on frequencies of concepts (words) co-occurrences in the pages indexed by search engines. Our experiments represent the better performance of the proposed approach in comparison with established trust evaluation approaches. The suggested approach can enhance efficiency of any solution which models semantic relations by ontology tree.*

## 1. Introduction

Trust is a critical concept in mutual collaboration in dynamic e-commerce systems. It is defined as a particular level of subjective probability using which, an agent assesses it and another agent will perform a particular action before it can monitor such action [1]. In the context of e-commerce systems, the actions are the e-commerce transactions. The trusting agent is called the trust or entity, and the trusted agent is called the trustee entity.

To evaluate the trustee's trustworthiness for a certain trust scope, context attributes is one of the two kinds of input analyzed by trust or [2]. Context attributes represent contextual information that the trust or requires in order to complete the evaluation of the trustee's trustworthiness. As a formal definition, context is any information that can be used to characterize the situation of an entity [3]. Context value for all the contexts may not be available. So, it is essential to have a mechanism for evaluating the unavailable trust value of certain context, using the available trust value of another context. It can be done in many different ways such as multiplying the trust value of the trustee in the available context into the similarity between available and unavailable contexts. As a result, computing the similarity between two contexts is crucial for trust evaluation in e-commerce systems.

There are many researches which attempted to compute unknown trust value of certain context, using the known trust value of another context. A significant portion of the researches utilize ontology trees for context modeling such as [4], [5]. These researches often exploit node distance to compute similarity. There is an underlying assumption in this exploitation: each two adjacent nodes have equal semantic distance or granularity of nodes in each level is identical. This underlying assumption is not true in most of the trees and it deteriorates the performance of trust evaluation. This research attempts to transcend this limitation by offering a novel weighted ontology tree, which is independent of the tree's structure. Our experiments on real data extracted context from Epinions.com shows that weighting trees improves the performance of trust evaluation.

The remainder of this paper is organized as follows. Section 2 surveys related works in context modeling and computing similarity between the contexts. In Section 3, essential materials for the proposed method are discussed in two subsection, similarity computation and ontology tree construction. Our suggested model is described in Section 4. Sections 5 and 6 are related to experimental setup and results, followed by a conclusion in Section 7.

## 2. Related Work

There are several previous works which aim to compute the mentioned similarity between to context in trust evaluation. To do so, in all of the researches first they used a model for context representation and then they introduced a method for computing similarity between the contexts. Therefore, we split this section according to these two steps.

### 2.1 Context Modeling

In order to compute the similarity between two contexts, the first step is to model the context which is known as context representation or context modeling. Any approach is used for the context modeling results different types of the similarity computation. Three popular types of these approaches are: ontology tree, key word based modeling and task based modeling [6].Of course, there are several other approaches which can be used in context modeling but they are not as popular as the above approaches. Strang et al. have a survey on these approaches [7].

#### 2.1.1 Ontology tree

Ontology tree is referred to the approach which the contexts are represented in a context ontology tree hierarchical structure. Each node in this tree represents a context and is split into two lower level contexts and the low level contexts are sub-context of the node. For example, Figure 1 shows ontology tree for network context and its sub-contexts [2].

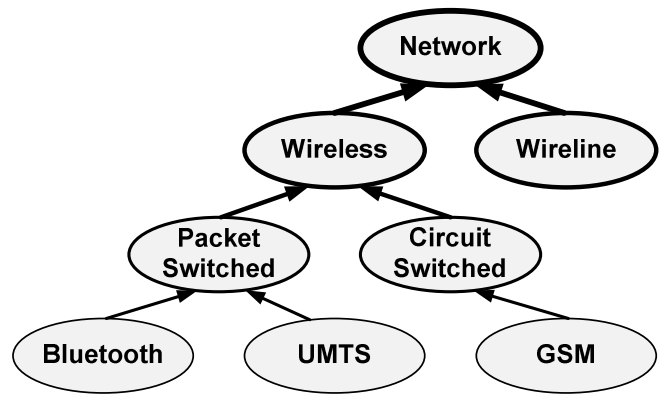

Figure 1.Example of an ontology tree

In [4] they make use of an ontology tree of services using DAML-S6, where each node in the tree representing a type of service.  Using ontology tree for representing game application running on a gaming device is another work which is done by [2].Here, a game application is composed by a game manager component (GM) and by one game scenario component (GS). In [8] they introduced a belief-theoretic reputation estimation model for multi-context communities. They employed an ontology tree to show consumer experience reports and beliefs about various products of a website (i.e. Epinion.com).

One of the limitations of these ontology tree approaches is that the tree may be constructed unfairly or granularity inconsistent. In particular, on branch of a node may be split generally while the other branch is split in more details which will be discussed in more details later. In this paper we mainly focus on this approach and introduce a method to overcome its limitations.

### 2.1.2   Keyword Based Modeling

Second common approach for context representation is using a combination of keywords to show a context. Each keyword is referred to a different context and by ensemble the keywords the result collection is a context. For example in all the papers there is a keyword section which introduces the main concepts which the paper has been written around it. Our paper keywords are: Trust, Context, Weighted Similarity, and Ontology. In [9] they used this approach for context representation. They considered a file-server application having three types of services (i.e., contexts): upload PDF File with keywords {*write, pdf, file*}, upload DOC File with keywords {*write, doc, file*}, login with keywords {*LoginInfo, userName, passWD*}.

The main advantage of this approach is its simplicity. Contrary to the previous approach, there is no need to perform any preprocessing to construct a tree and it can be applicable in any context. But their disadvantage is their limitation in extension. There are some situations where it is not possible to specify the context by using some simple labels.

### 2.1.3 Task Based Modeling

The third approach is more applied method and is built on tasks. Suppose that we are working on a certain environment with certain jobs. In such a situation, the collection of tasks which can be done is limited and will not be exceeded from a certain threshold. Therefore, in such cases each task can be considered as a context. Here, each task is composed of several sub-tasks which are knowntask's aspect or task's attribute. An aspect is the smallest element of a task which describes a special attribute of it. In [6] they worked on several tasks such as: "Tom is wondering about trusting Bob to guide him in London when it is stormy". Here, the task is model as: Location: London, Weather: stormy, Subject: guide. As it may be guessed the task's aspects are: Location, Weather and Subject. This approach is also employed in other researches such as [4, 9, 10].

This kind of context modeling cannot be used in general and is limited to specific cases. In particular, when we are facing with a situation where the collection of possible tasks is limited, the tasked based modeling can be an appropriate solution.

There are several other approaches which can be used in context modeling but they are not as popular as the above approaches. For more study the different approaches can be found in [9].

## 2.2 Computing Similarity between the Contexts

After identification of a model to represent a context, the next step is to specify a method to compute similarity between the contexts. In this section the goal is to introduce these methods which have been used in previous researches.

In [4] similarity between two contexts is computed by the distance between to node in the context's ontology tree:

$$Simialrity\ (S1, S2) = \frac{1}{Distance(S1, S2)} \qquad (1)$$

Here, the distance of two nodes is defined as the least number of intermediate nodes for one node to traverse to another node. For example, in Figure 2 which shows services ontology tree, service s1 and s2 has a distance of 3.

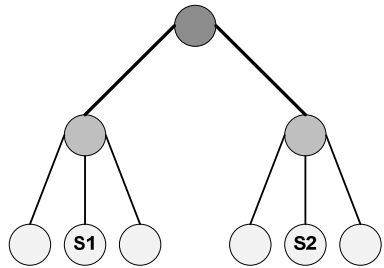

Figure 2.Services in a context ontology tree [4]

[2] introduced another similarity computation method for contexts which are represented in an ontology tree. Here, the similarity between two nodes is calculated as the ratio between the

number of shared nodes from the source node and the sink node to the root node, and the total number of nodes from the source and the sink to the root node. For example in Figure 2 *s*1 and *s*2 has a distance of 3/5.

[9] considered any context as a set of keywords and they computed the similarity between two contexts by using the set theory. Here, the similarity between two contexts, $S_i$ and $S_j$, with their individual keywords sets, $K(S_i)$ and $K(S_j)$, is defined as the ratio between the set's intersect and the set's union:

$$Simialrity\left(S_i, S_j\right) = \frac{K(S_i) \cap K(S_j)}{K(S_i) \cup K(S_j)} \qquad (2)$$

As it was elaborated, one approach for context representation is considering a context as task. In [11] the similarity *D(S1,S2)* between two tasks s1 and s2 is obtained from the comparison of the task attributes.

$$Simialrity\left(S_i, S_j\right) = 1 - \frac{1}{n}\sum_{l=1}^{n}\left|S_{i,l} - S_{j,l}\right| \qquad (3)$$

where n is the number of task attributes, $S_{i,l}$ is the *l*-th attribute of task $S_i$, and $S_{j,l}$ is the *l*-th attribute of task $S_j$.

In [6] in order to measure similarity among contexts, they used the idea of the bipartite SimRank which is an extension of the basic SimRank algorithm [12] to bipartite domains consisting of two types of objects. Such domains are naturally modeled as graphs, with nodes representing objects and edges representing relationships. Here, they formed a graph with contexts and aspects as nodes. In this graph each context points to their aspects (Figure 3). The recursive intuition behind this algorithm is that in many domains, similar objects are related to similar objects. More precisely, contexts A and B are similar if they are related to aspects b and c, respectively, and b and c are themselves similar. The base case is that aspects are similar to themselves.

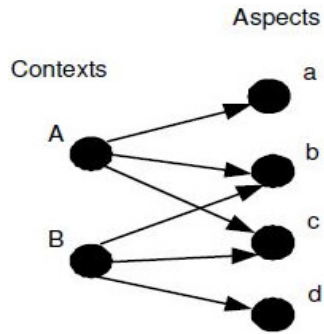

Figure 3.Graph model of context in [6]

## 3. Methods and Materials

The proposed solution utilizes concept similarity computation and ontology tree as two base materials. In each of these areas, there is rich literature representing the importance of the research topic. We choose the proper method based on our requirements and the experiment result comparison of the methods. In this section we describe used methods and relating subjects. Next two subsections introduce our method on concept similarity computation and semantic hierarchical structure respectively.

## 3.1 Normalized Similarity Score

To develop the ability of text understanding for computers, two major approaches are adopted so far: using expert-created semantic structure and automatically extracting semantic relation from human-written text. Based on the first approach, several large and long-term projects are established such as Cyc [13] and WordNet [14]. These projects try to establish semantic web of vast variety of concepts, which comes at enormous effort and cost. Despite of these efforts by knowledgeable human experts, this approach has a significant limitation: In comparison with available information on the web, the total entered information is limited [15]. Covering this limitation, the second approach is developed in the recent years. The new approach utilizes the large public available user-generated data on the web to achieve semantic relations which is accessible on public available search engines. Most of the methods based on the second approach employ aggregate page-count estimates of search-queries to extract semantic relations. In this research, we use the second approach for concept similarity computation. Poor quality of the first approach in our evaluations directs us to the second approach.

Concept similarity can be determined out of co-occurred words' frequency in articles automatically. Normalized Similarity Score (NSS) uses these frequencies to measure semantic relatedness between words [16]. This score is derived from Normal Google Distance (NGD) [15]. In order to utilize NGD as a relatedness measure -rather than a distance measure-Lindsey converts NGD scores into similarity scores by subtracting NGD from the its maximum score. Therefore NSS computes the relatedness between two terms a and b as follows:

$$NSS(a, b) = 1 - NGD(a, b) \qquad (4)$$

NGD measures the distance between two terms by the symmetric conditional probability of their co-occurrences [17]. It means that NGD assumes that the probability of word $x$ co-occurring along with word $y$ is high when the similarity between their concepts is "near" to each other and vice versa. NGD is formulated as following equation:

$$NGD(x, y) = \frac{\max(\log f(x), \log f(y)) - \log f(x, y)}{\log M - \min(\log f(x), \log f(y))} \qquad (5)$$

where $f(x)$ is the number of times a search engine hits for the search term x; $f(x, y)$ is the number of times this search engine hits both of $x$ and $y$ simultaneously; and M is the total number of pages that can potentially be retrieved in search engine (e.g., Google can potentially retrieve around 10 billion pages) [18]. Originally, NGD was developed for using by Google search engine; nevertheless it is applicable in other search engines as well. In the present research Bing is selected as a search engine due to its better performance.

## 3.2 Ontology tree construction

There are two practical approaches for constructing ontology trees: utilizing Word Net hierarchical semantic structure and extracting ontology tree from e-commerce website categories. Each of these approaches suffers from several problems. To alleviate the problems, combining these two approaches is one of possible solutions. In the current research, this solution is used. The rest of this subsection introduces these approaches and details the strength and weakness of them.

### 3.2.1 Using the Word Net Ontology tree

Word Net [7, 8] is a hierarchically organized lexical system motivated by theory of psycholinguistics that was developed at Princeton University in the 1990s. As a conventional online dictionary, Word Net lists alphabetically concepts important to a particular subject along with explanation. The major advantage of Word Net is linking the words based on semantic relations between their meanings [21]. The most frequently encoded semantic relation among synsets is the super-subordinate relation i.e. hypernym-hyponym. This relation  links more general synset to the specific ones. Hypernym represents *is-a* relationship among the words. Contrarily, hyponym is *inverse-is-a* relationship. As an example, {*digitalcamera#1*} is a hyponym for {*camera#1*} and a hyponym for {*webcam#1*}. Figure 4 depicts the hypernyms tree for webcam. Hypernym-hyponym relation can be utilized to extract semantic hierarchy structure (or ontology tree). But, another problem exists yet. It is possible that a word have multiple parents in at the same level of hierarchy. To face this issue, we select one of the more significant parents based on the meaning of them. For example, however {*camera*} has two hypernym: {*photographic equipment*} and {*television equipment, video equipment*}, we use {*photographic equipment*} for ontology tree extraction. Because our mean by the word *camera* is a device for take photograph.

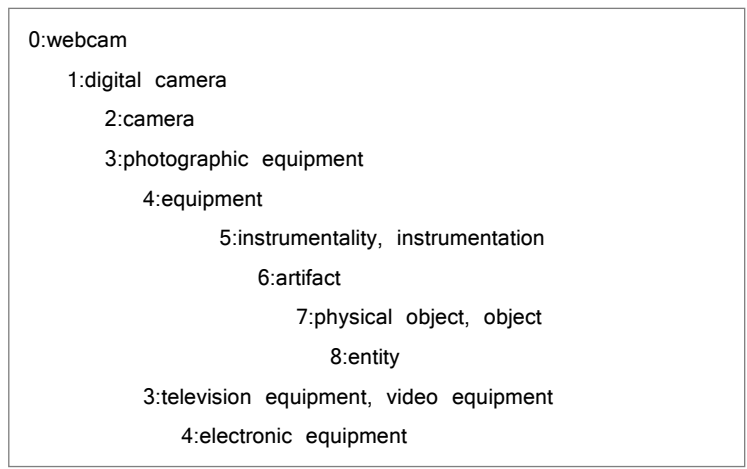

Figure 4. Hypernyms tree of "webcam"

Using WordNet hierarchical semantic structure is widespread in research projects; however, this structure is not applicable in real applications for a few reasons. First, concepts are categorized by their semantics rather than their applications. It makes two close concepts to become far from each other in real world context. For example, while in real stores both monitor and monitor cleaner are in the same category, in a semantic tree they are not. Second, WordNet is unfair. It means that the abstraction ratio in the tree levels is not equal for all concepts. This problem makes similar words to be at different depths in the ontology tree. More clarification regarding to the mentioned problems is shown in Figure 5. This figure displays the positions of three similar words in WordNet tree: Mouse, Keyboard and Laptop. As seen, while all electronic stores categorize "mouse" and "keyboard" in the same level, WordNet does not. In addition, exhibited distance and depth difference between "keyboard" and "Laptop" does not seem to be true.

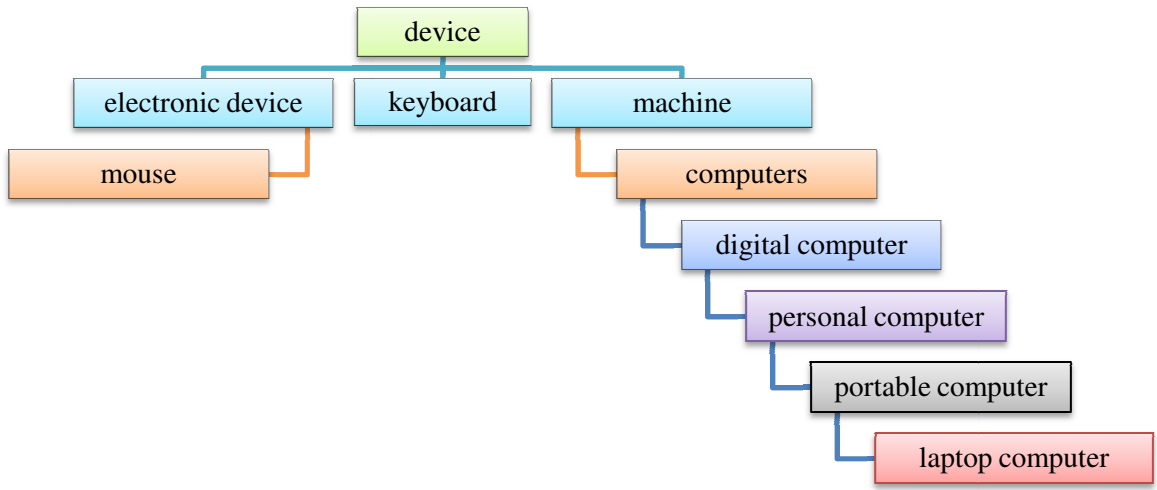

Figure 5. Semantic granularity is not equal all over the WordNet. Depth of Mouse, Keyboard and Laptop in WordNet hierarchical semantic structure does not make sense.

### 3.2.2 Using the ontology tree extracted from website categories

Extracting ontology tree from product categories of e-commerce websites is another approach to overcome the limitation of the Word Net tree. However, there is not any publicly available dataset based on this approach. Several website such as Netflix have flat categories, while hierarchical structure is necessary for our purpose. Another essential requirement of ontology tree is granularity consistency i. e. each hierarchy level of tree should be almost in same semantic detail level. Among directory and ecommerce websites (such as Yahoo Dir., and Amazon) eBay comparatively satisfy this requirement more preferable. Moreovere Bay has another benefit: its ontology tree includes comprehensive range of shopping concepts, since it sells various kinds of goods. Figure 5 depicts the full ontology tree extracted from eBay.

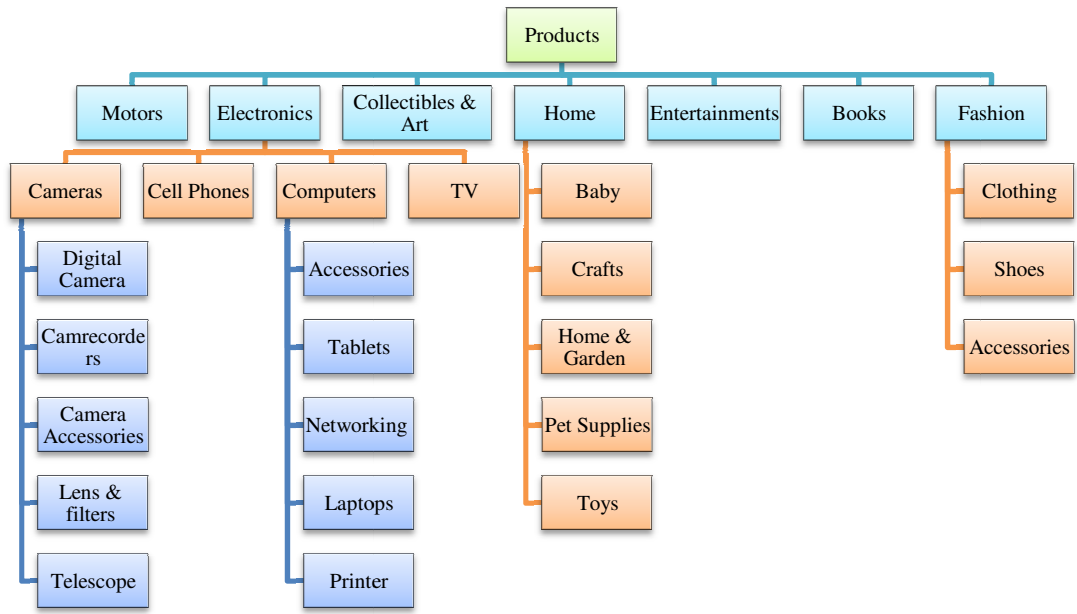

Figure 6. Ontology tree which is extracted from eBay categories

Despite the mentioned advantage of eBay ontology tree, it is far from a mature ontology tree yet. This tree covers a few contexts in comparison with WordNet. In addition, the contexts are categorized by their applications rather than their semantics, contrast to WordNet. It makes two distant concepts to become adjacent in the ontology tree. For example contrary to common sense, in Figure 6 "Home" is the parent (more general concept) of "Baby".

As aforementioned, ontology tree of WordNet and eBay is on the two end of a semantic-applied spectrum. WordNet is completely semantic, while eBay is applied. Each of them causes a specific difficulty. A reasonable approach to reduce difficulty is combining two previous approaches. Hence we prefer combination approach and we construct several ontology subtrees which are figured in section 6.

## 4.  Proposed Approach

In this paper, we attempted to show an advanced ontology tree for context representation overcoming the limitation of the previous trees. Afterward, we detail an enhanced method for computing the similarity between two contexts based on the proposed tree. To do so, in the section first, we reveal the limitation of the previous methods and then the proposed enhanced solution will be shown.

### 4.1  Limitation of ontology context modeling

In section 2, we elaborated three approaches for context modeling and pointed out their limitations. Among these approaches the most popular one is the context modeling using ontology tree. As discussed before, the most important limitation of this approach is that the tree may be constructed unfairly. In particular, one branch of a node may be split abstractly while the other branch is split in more details. In other words, this tree is granularity inconsistent. The limitation is illustrated in the following (Figure 7):

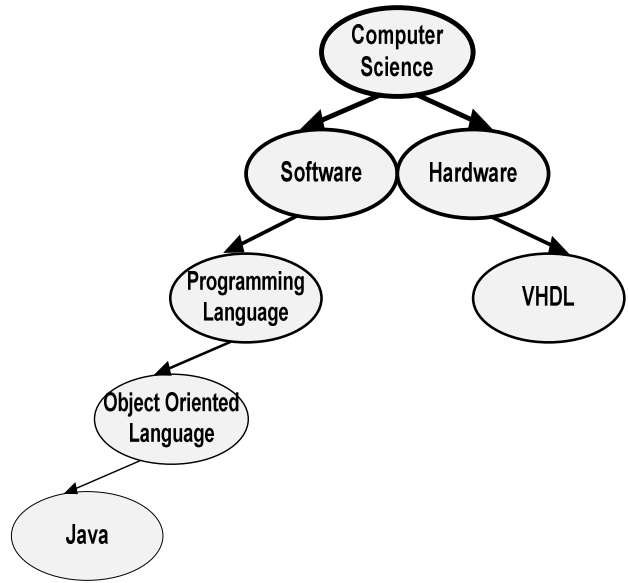

Figure 7.An example of unfair constructed ontology tree for the computer science concepts

As shown, computer science is split to software and hardware. Afterward, the hardware node is split to VHDL programming language while the software is split to programming language; afterward object oriented language and finally the java programming language. As seen, in the above tree VHDL and java are both a programming language in hardware and software context but their distribution is not equitable. In particular, the distance between hardware and VHDL is an edge while the distance between software and java is three edges and so it is not an equitable distribution. Therefore, the VHDL node should be split into more nodes in order to have a fairly constructed ontology tree. As it is clear, this unfair construction of the ontology tree causes several problem in the context's similarity computation methods which are based on these ontology trees.

## 4.2 Context modeling based on weighted ontology tree

In favor of overcome to the described limitation, we suggest to use a weighted ontology tree instead of the traditional trees. Edges weights in this tree represent the similarity between their corresponding nodes. To clarify the issue, it is illustrated by the Figure 8. By specifying the similarity between the nodes of an edge, the distance between any two arbitrary nodes can be specified more equitable. Therefore, the total distance between hardware and VHDL is equal to the total distance between software and Java (i.e. 14). The reason is that, despite of splitting the software node in more details the distance between the split branches is not much and so both total distances become equitable.

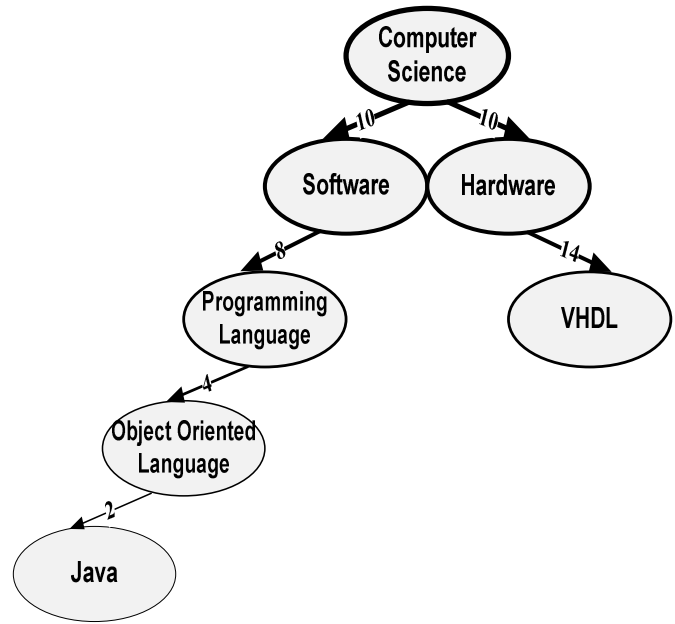

Figure 8.Fairly constructed ontology tree for the computer science concepts

In order to implement the above solution it is needed to construct a weighted ontology tree. To do so, first we need to have a method for computing the similarity between the nodes of an edge as their weighted distance. To achieve this, we use the Normalized Similarity Score (NSS) method defined in subsection 3.1. This method is based on frequencies of concepts (words) co-occurrences in the pages indexed by search engines. Here, each context is a concept, which has its own meaning in the dictionaries. The ontology tree's edge will be labeled by the similarity between its two ends nodes. Afterward, the similarity between any two arbitrary contexts can be computed by multiplying the edges weight on the path between them in their ontology tree. For instance in Figure 9, multiplying w1, w2, w3 and w4 results the similarity between S1 and S2. Thus, we can formulate the similarity between any two arbitrary contexts $C_i$ and $C_j$ as:

$$Similarity\left(C_i, C_j\right) = \frac{1}{\prod_{k \in Path(S_i, S_j)} w_k} \tag{6}$$

where the $S_i$ and $S_j$ are the related node of $C_i$ and $C_j$ in the ontology tree. In addition, $w_k$ denotes the weight of edges in the unique path between $S_i$ and $S_j$. Using the above method, distance between two nodes and the edges' weights have impact on similarity simultaneously.

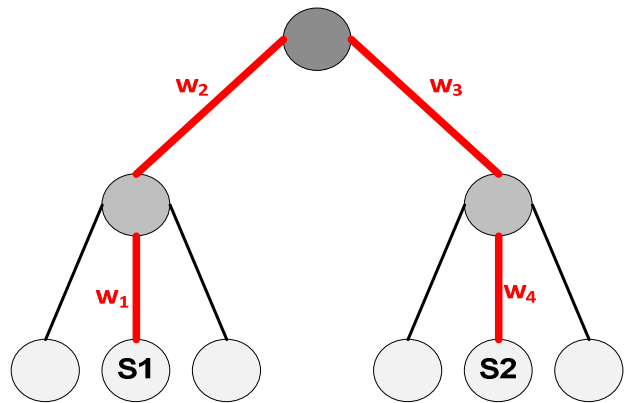

Figure 9.Services in a context ontology tree

## 5.  Experimental Setup

Over the last decade, publications on computational trust model have significantly increased. However, these researches seldom have evaluation on real data. Among the research that evaluated their model, most of them have used simulation techniques using stochastic generated data. Therefore, evaluation of trust models with real data is still required to investigate their practical consequences. In the present research, we aim to evaluate our proposed method on a real data set. To do so, two notable issues should be carefully considered:

1.  There is no public dataset available on trust area including context of each transaction (based on our literature review). Regarding available datasets such as Epinions, transactions are not linked with their related real record to find their context; therefore, data should be collected from scratch.
2.  There is not any standard process to evaluate the results in context-aware trust modeling, thus a process for evaluation of the proposed method should be suggested. The process should depict the difference between the accuracy of the trust modeling in the simple and weighted ontology.

To cover the above concerns, we considered several solutions, which are studied in the following subsections.

### 5.1. Data collection

We extract our data set from Epinions.com. Epinions is a review website where ordinary users can assign rating and write reviews about product and seller. Also they can assign a trust rating representing helpfulness, to reviewers. Users can access to recommendations, criticisms, and reviews for products; however, only registered users are permitted to participate in rating a product or writing reviews at Epinions [22].

To collect data, various popular e-commerce sites such as eBay, Amazon, and Epinions were investigated. Each of these websites has its own limitations to be used in our evaluation. For instance, eBay offers the average rating of all customers (reputation) on each seller, whereas each transaction rating is needed for our study, because we should determine the context of each transaction as well as its corresponding rating. Regarding Amazon, since buyers only rate products (not sellers), computing trust of a seller in other contexts is impossible. Contrary to eBay and Amazon, Epinions.com can be a suitable choice for our purpose, despite of its shortcomings. In Epinions users can rate seller as well as products. Moreover it offers the ratings on seller separately. As a result, Epinions does not have the aforementioned limitations of eBay and Amazon.

Data collection on Epinions encounters with a two major challenges. First, what kind of reviews should be collected? And second, how can we collect these kinds of reviews? This subsection elaborates on these challenges, and how we finally collected an appropriate dataset for our experiments.

First, only trustee who has multiple contexts is appropriate for our experiments. Therefore, products' ratings (most of the Epinions reviews) are inappropriate for our requirement, since a product is not definable as a trustee and does not have multiple contexts. Regardless of the ratings on products, one category of Epinions.com remains including multi-context data: "Online Store and Services". In this category, users can rate e-stores (not product) and write reviews about them. As these stores sell various kinds of products, the corresponding user reviews could be in different contexts. Thus, we collect the required data from reviews of "Online Store and Services" category. Also, among these reviews, only those regarding a unique context are usable. To be exact, overall reviews without focusing on any context or reviews deal with multiple contexts (concerning purchasing numerous products) are not suitable for our purpose.

Second, context of each rating is not clearly available in Epinions, and so, automatic gathering of context data by conventional web scrapper is impossible. Therefore, in order to identify the context of a review, its text should be studied by human. For instance, if in a review, a user has commented on the quality of a Lego, bought for her child, the toy context should be assigned to this review. Moreover, in addition to difficulty of context identification, appropriate sample size is another limitation of data collection. For each seller, we require at least two contexts including approximately 30 ratings on each context, while most of the sellers do not have as many ratings.

| Context | Rate | Date | Description | Link |
|---|---|---|---|---|
| Laptop | 4 | 12-Aug-09 | Macbook Pro | http://www.epinions.com/content_48103650 |
| Laptop | 5 | 27-Apr-08 | laptop computer | http://www.epinions.com/content_42792547 |
| Laptop | 1 | 5-Jul-03 | IBM laptop T23 | http://www.epinions.com/content_98326515 |
| Laptop | 3 | 17-Dec-03 | toshiba laptop | http://www.epinions.com/content_12214970 |
| Laptop | 1 | 18-Jul-07 | Apple Laptop | http://www.epinions.com/content_39247810 |
| Laptop | 5 | 25-Mar-04 | Laptop | http://www.epinions.com/content_13444307 |
| Laptop | 1 | 6-Nov-06 | laptop | http://www.epinions.com/content_28433321 |
| Laptop | 1 | 2-Apr-02 | Rip OFF | http://www.epinions.com/content_61829582 |
| Laptop | 4 | 9-Nov-03 | desktop | http://www.epinions.com/content_11805095 |
| Laptop | 4 | 1-Aug-01 | Be Careful New | http://www.epinions.com/content_31329848 |

Figure 10. Overview of fields that collected

The process of data collection on Epinions regarding the above challenges is as follows. We skim thousands of reviews and ignore reviews that are unrelated, without any specific context or with multiple contexts. The ratings' data of the remained reviews is collected according to their seller-context classification separately. Finally, sellers which do not have at least two contexts including approximately 30 ratings are removed from data. Despite of the mentioned challenges, we gathered ratings data off our sellers in different contexts supporting our experiments. These sellers are eBay, Overstock, Beach Camera and Amazon. Figure 10 depicts some part of the collected data in Laptop context. This data consist of five fields: context, rate, rating date, description and URL (i.e., link to the source of the review).

## 5.2. Evaluation Criteria

As aforementioned, final goal of this paper is to predict the trust value of a certain user in an unknown context, based on their trust value in a known context. To evaluate the accuracy of our prediction, an evaluation measure introduced by Liu et al. [4] is utilized. This measure calculates outcome error from formula (7). This formula is a kind of "Prediction Error" type. This type of error calculation is one of the most widespread perform anceevaluation criteria exploited in several other papers on trust models [24, 25, 26] .According to these papers, prediction error of a trust evaluation model can be computed as follow:

$$Error\ percentage = \frac{Predicted_{Rate} - Real_{Rate}}{5} \times 100 \tag{7}$$

Where *Predicted_Rate* is the predicted trust value of the trust evaluation model, and *Real_Rate* is the actual trust value.

# 6. Experimental Results

To evaluate our proposed method, at first, it is applied on several subtrees extracted from the base ontologytree (see Figure 6), elaborated in the previous section. Second, all the tree edges are weighted using Normalized Similarity Score detailed in subsection 3.1. The resultedweighted subtrees are shown in Figure 11.

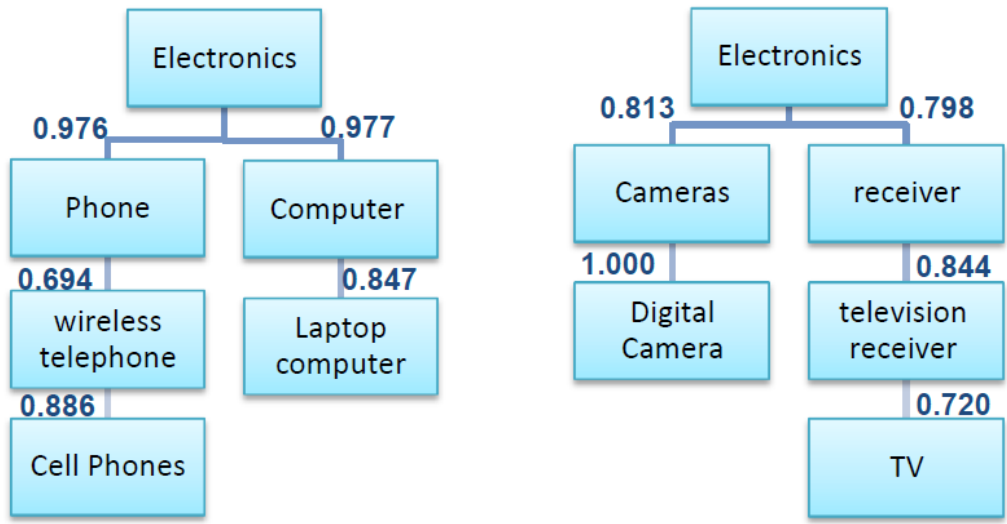

A) Subtree between Cell-phone and Laptop for eBay data    B) Subtree between Digital-Cam and TV for Beach Camera data

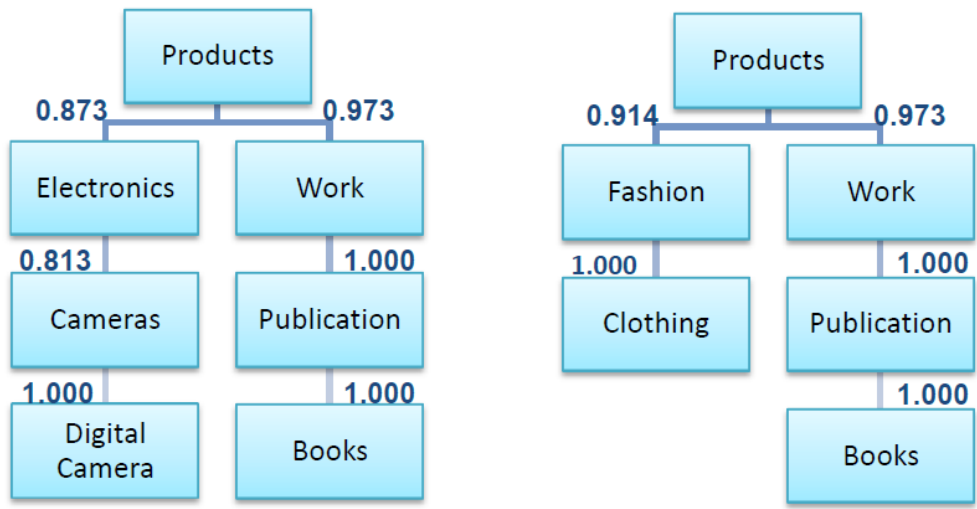

D) Subtree between Digital-Cam and Book for Amazon data    E) Subtree between Clothing and Book for Overstock data

Figure 11.Subtrees constructed to evaluate proposed method

In the third step of experiment process, the trust evaluation criteria, described in subsection 5.2, is applied to both our proposed method and the Liu et al. similarity computation method [4] formulated in equation (1), in order to compare weighted and un weighted similarity computation methods respectively. These methods try to predict trust in an unknown context using a known context. As Figure 10 compares the error of these predictions, our proposed method outperforms the prediction results. The reason for deficiency of the un weighted method is its static approach on similarity computation. As mentioned in subsection 4.1, this method considers only the path length between two concepts in the ontology tree. As a result, when the tree is unfair, the path length between two concepts is not remarkable which represents the similarity lower than real value. For example, in left bottom subtree of Figure 5, adding "work" and "publication" nodes between "Books" and "Products" increases granularity of this branch. Accordingly, the subtree become sun fair and the similarity will be decreased to 0.2 and the trust value on the target context is predicted with less accuracy. On the contrary, our proposed method decreases this drawback using semantic similarity of each parent and child nodes.

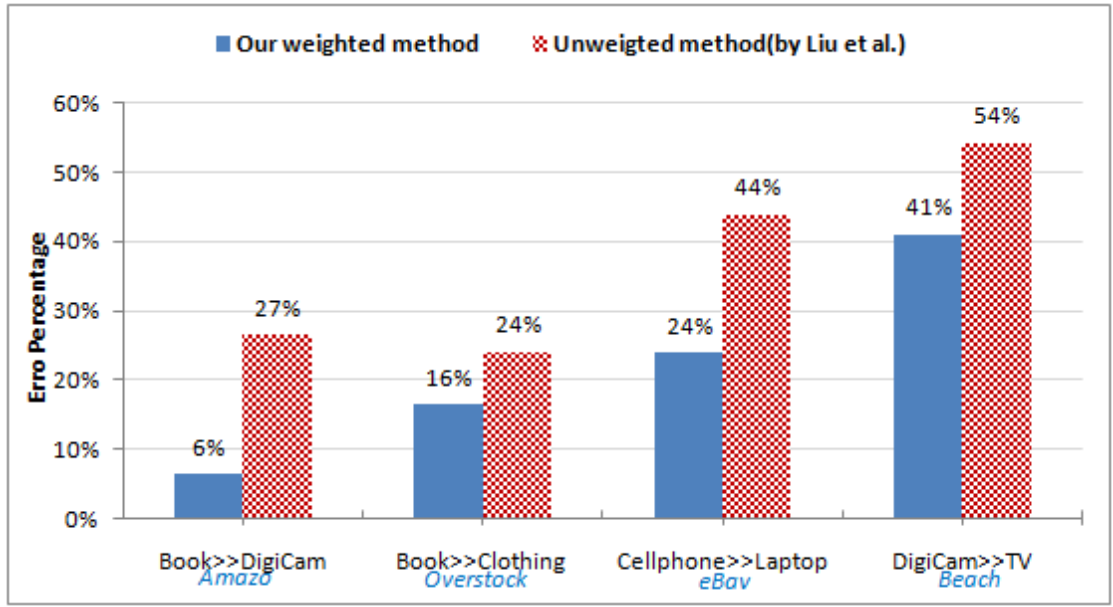

Figure 12.Comparison between prediction error rate of the proposed method and un weighted method [4] on real data

The most performance improvement of Figure 12 is occurred on eBay and Amazon cases. Related subtrees of these cases (see Figure 11) explain the reason. These subtrees are fairer compared to others. In eBay case adding "phone" causes semantic fair of the two branches, while adding "Cell phone" increases granularity of the left branch. Also, in Amazon tree "Book" and "Digital Camera" are in the same level of tree according to our expectation, while in other trees leaves have dissimilar levels.

Until know, error of proposed method was compared to unweighted method error proportionally, whereas absolute error pattern of our method is another substantial issue, shown on Figure 12 results. This figure exhibits that the least prediction error achieved by Amazon. The reason of this achievement can be due to Amazon's expertise in book context, which has gained popularity for the electronic market. Accordingly, its trust value on book is higher than other contexts. Therefore, our method can predict Amazon trust on "Digital Camera" context accurately. In figure 14 the relative expertise between two contexts is defined as "rate difference". It signifies the absolute difference between real trusts rates on two contexts. Figure 13 represents the relation between the rate difference and the proposed method error. The more prediction error increases, the less rate difference decreases. It indicates that expertise has an important influence on prediction performance, and the high Pearson correlation coefficient between these variables (about -0.95) confirms this claim.

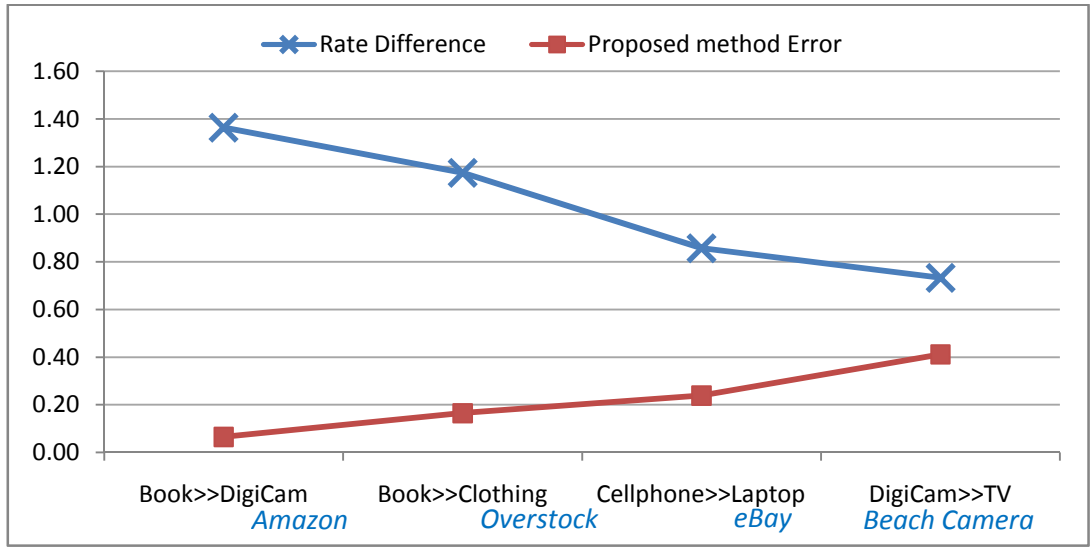

Figure 13. Relation between real trust rate difference and proposed method error

## 7. Conclusion

This research transcended a limitation of previous ontology tree context modeling to improve context similarity measurement. An important limitation of context modeling using ontology tree is that the tree may be constructed unfairly or granularity inconsistent. In other words, the semantic similarity of each two adjacent nodes is unequal in the ontology tree. The proposed approach overcomes this limitation by weighting edges based on their semantic similarity. Weight of each edge is computed based on Normalized Similarity Score (NSS) method. This method is based on frequencies of concepts (words) co-occurrences in the pages indexed by search engines. Using the proposed approach, trust value prediction of a certain user in an unknown context, based on their trust value in a known context becomes more accurate. Thus, this approach can be implemented in a wide range of web applications from a small business environment to a large market-place such as electronic shopping systems.

To test the success of the proposed approach, we collect customer reviews about four e-commerce sellers in Epinions.com. For each seller reviews of at least two contexts were collected. It is assumed that trust value in a context is known and the other is unknown. We compute trust value in the unknown context from the known context. We perform this computation twice, once with weighted ontology tree and once with unweighted. The difference between these two results show the performance of the proposed approach compared with previous approach.

Our experimental results showed the performance of the proposed approach over unweighted ontology tree. The prediction error of trust evaluation with weighted ontology tree is 8 to 21 percent lower than unweighted one under different scenarios. As tree become fairer after weighting, the performance improvement becomes more obvious. In addition to relative error (performance compared with previous approach), absolute value of error also follows a certain pattern. The absolute error of the suggested approach was less when we utilize trust value of the context which trustee is expert on that context. If we define expertness as difference between real ratings of known and unknown context, expertness has high negative correlation with absolute error. Amazon is an example of this fact. Amazon is an expert website in the context of book. Accordingly, predicting the trust values of Amazon in other contexts based on book context is more accurate. It is worth noting that this feature is often useful. Most of the times we know the trust value in popular context of a seller and we require predicting trust values of other contexts.

The novelty of the current research relies on two major facts. First, the proposed approach improved the performance of trust evaluation in unknown contexts. Second, we collect a real trust data set including context information with considerable effort. This is done while previous researches on context trust evaluation either do not asses their models or use simulation for test. Obviously, the result on real data is more creditable than simulation. In addition to the mentioned contributions, this study has other contributions such as: the procedure of evaluating the proposed approach, the method of ontology tree construction, and using automatically extracting semantic relation from human-written text for weighting ontology tree.

As a future work, the proposed approach should be evaluated on larger data set and other application (instead e-commerce). Another option for continuing this research is comparing the performance of weighted and un weighted ontology tree outside the area of trust and reputation. Furthermore, suggesting a method for expertness measurement enables us to estimate the performance of trust evaluation. Another avenue of exploration is to extend suggested similarity computation method to normalize the edges' weight in each problem. It can be embedded to our model with configurable parameters.